\documentclass[12pt]{article}
\usepackage{graphicx}

\usepackage{amsfonts}
\usepackage{amssymb}
\usepackage{amsthm}
\usepackage{amsmath}
\begin{document}

\begin{titlepage}
%\begin{flushright}
%hep-ph/0506144{\hskip.5cm}
%\end{flushright}

\centerline{ \hfill hep-th/0506144 } \vspace{3.5cm}

\begin{centering}
%\vspace{.5in}
%
{\bf \Large
Standard and Non-Standard Extra Matter for Non-Supersymmetric
 Unification}\\
\vspace{1.5cm}
{\large Alex Kehagias and N.D. Tracas} \\
\vspace{.6cm}
{\it Physics Department, National Technical University,\\
       Athens 157 73 , Greece}\\
\end{centering}
\vspace{2cm}

\abstract{\noindent } \noindent {
%\normalsize
We perform a general 1-loop analysis by adding  extra matter to
the SM content, as well as by allowing a non-standard $U(1)_Y$
normalization in order to achieve unification.
We find numerous solutions with $U(1)_Y$ only charged extra matter
and unification scale of the order of $M_U=10^{16}$ {\rm GeV},
or with $SU(2)$ charged extra matter with lower $M_U$. We  next
identify the SM extra matter as originating from the breaking of
$SO(10)$ either through the Pati-Salam group $SU(4)\times
SU(2)_L\times SU(2)_R$ or the flipped $SU(5)$. In both cases,
unification can be achieved with a rather minimal extra content.
Contrary to the one-step $SU(5)$ unification, in the two-step
$SO(10)$ unification we are discussing, coulored extra matter is
not necessary.
Finally, we discuss a split-supersymmetry like
case, where extra matter is added to the split supersymmetric
spectrum in order to attain 1-loop unification. }
\vfill \hrule
width 6.7cm
\begin{flushleft}
June 2005
\end{flushleft}
\end{titlepage}

\noindent
\section{\large{Introduction}}

 Understanding the electroweak
symmetry breaking is one of the most important problems in high
energy physics. Ultimate connected with this is the new physics
expected to be found in the forthcoming experiments. Many ideas
have been proposed such as   supersymmetry, extra dimensions
e.t.c. Among these, weak scale supersymmetry is the most popular
one, as among others, it gives answers for the gauge hierarchy
problem, the origin of the electroweak symmetry breaking as well
as it provides dark matter candidates. Moreover, it
 has a concrete prediction, namely, gauge
coupling unification around $M_U\sim 10^{16}{\rm GeV}$. This fact
has been considered as supporting evidence for both gauge
unification and supersymmetry, pointing towards the idea that
supersymmetry is indeed realized in nature.

 However, one may want to extrapolate SM  beyond the TeV
threshold, demanding the celebrated features of supersymmetry,
like gauge coupling unification and dark matter candidates, but
without supersymmetry itself. In this case, we are facing the naturalness problem of
the scalar sector, which makes this extrapolation above the {\rm
TeV} scale questionable. This is much like the cosmological
constant and the naturalness problem connected with it, which however
is ignored in all practical calculations.
Similarly for the case at hand,  one may assume that
there is a mechanism which makes the TeV scale harmless and SM can
be followed up to GUT energy scales.
This line of though has recently be proposed in~%
\cite{AD}. In this scenario, there is no low-scale supersymmetry
but rather supersymmetry is realized at some intermediate scale
$M_S$
and expected gauge unification still at the GUT scale $M_U$. This
is an alternative to the  MSSM
with low-energy supersymmetry and it is known as split supersymmetry~%
\cite{AD},\cite{Giudice:2004tc},\cite{AD1}. In the latter, all superpartners
are pushed to the
 $M_S$, while only gauginos and Higgsinos remain at the weak scale. Apart from the latter,
the fine-tuned Higgs as well as the fermions, which are protected
by chiral symmetry, remains also near the electroweak scale.
Clearly, this proposal violates the naturalness requirement for
the Higgs mass and it seems to contradict the reason of
introducing supersymmetry in first place. One may still assume
that there exist still an unknown mechanism which permit the Higgs
mass to remain around the weak scale. We recall at this point  the
cosmological constant problem mentioned above, where  consistent
calculations can be done ignoring the cosmological constant and
any mechanism connected with it. Recently, split supersymmetry was
shown to be realized also in string theory~\cite{AnD},\cite{KN}.

We recall that supersymmetry is a basic ingredient of string
theory and  if  supersymmetry is not realized at the TeV scale, it
is more natural  to be tight to the string scale. In particular,
it is very difficult to justify the requirement of a
supersymmetric spectrum as there is no reason for supersymmetry in
first place if we assume that supersymmetry plays no role in the
hierarchy problem. However, gauge coupling unification in the
supersymmetric SM  is an impressive aspect of supersymmetry and
one may wondered if it is possible to achieved it in
non-supersymmetric theories. The purpose of the present work is
exactly to check if gauge coupling unification may be achieved by
introducing extra matter (scalars and fermions) above the
electroweak scale. We first analyse the general case finding out
several irreps of extra matter which achieve gauge coupling
unification. We then, by considering a non-SUSY $SO(10)$ model, we
specify the extra matter needed in a 2-step unification:
$SO(10)\rightarrow G\rightarrow SU(3)\times SU(2)\times U_Y(1)$.
There are two possibilities for $G$, namely, (i) the Pati-Salam
$G=SU(4)\times SU(2)\times SU(2)$ \cite{Pati:1974yy} and (ii) the flipped $SU(5)$,
$G=SU(5)\times U(1)$~\cite{Barr:1981qv},\cite{Derendinger:1983aj},\cite{Antoniadis:1987dx}.
We will consider both cases here and we will
investigate one-loop gauge coupling unification in both models. We
will assume that new matter exists around {\rm TeV} range, which
contributes to the running of the gauge couplings above this
scale. The question we will try to answer is the form of this
extra matter for which gauge coupling unification is achieved.

We should mention that similar work have been done
in~\cite{Giudice:2004tc}. However, they have considered only
one-step unification where $SU(3)\times SU(2)\times U_Y(1)$ is
unified to $SU(5)$. We have confirmed their findings, and we have
found some new results in the case of a two-step unification,
i.e., in the case of an $SO(10)$ unification with a partial
intermediate unification of Pati-Salam or flipped $SU(5)$
unification. In the one-step $SU(5)$ unification, coloured
particles are necessary for achieving unification at high enough
$M_U$ for proton stability. In the two-step $SO(10)$ unification
we study here, unification may be achieved without
introducing coloured particles . In both cases, there are stable
dark matter candidates, while the splitting of irreps (the
double-triplet splitting in the case of $SU(5)$) is the same for
both one-step and two-step unification\footnote{
Extra matter in the form of fourth generation may also lead, under certain conditions,
to unification without SUSY\cite{Hung}%
}.

%\vspace{0.5cm}
\noindent
\section{\large{Gauge coupling unification}}
The one loop  gauge couplings running reads:
\begin{equation}\label{betas}
\alpha^{-1}_i=\alpha^{-1}_{i_0}-\frac{b_i}{2\pi}(t-t_0)-\frac{b^e_i}{2\pi}(t-t_0),
\quad\quad i=Y,2,3
\end{equation}
where $t$ is the logarithm of the energy, $\alpha_{i_0}$ is the
value of the coupling at the scale $M_0$, $b_i$ is the usual
$\beta$-function one-loop coefficients for the SM and $b^e_i$ is
the contribution from extra (non SM) matter. Assuming that at
scale $M_U$ the gauge couplings unify (i.e. $\alpha_i$ has the
same value for al $i$'s), we can write the following simple
relation between the differences of the $b$ coefficients:
\begin{equation}\label{dif}
\delta
b^e_{ij}=\delta\alpha^{-1}_{ij_0}\,\frac{2\pi}{t_U-t_0}\,-\delta
b_{ij}
\end{equation}
where $\delta b_{ij}=b_i-b_j$ and
$\delta\alpha^{-1}_{ij_0}=\alpha^{-1}_{i_0}-\alpha^{-1}_{j_0}$. In
case the normalisation of $\alpha_1$ is different from 1, i.e. at
$M_U$ we have the relation: $k/\alpha_Y=1/\alpha_2=1/\alpha_3$,
Eq.(\ref{dif}) holds with the following substitutions:
$b_Y(b^e_Y)$ should be changed to $kb_Y(kb^e_Y)$ and $\alpha_Y$ to
$\alpha_Y/k$.

If we further assume that there are no extra particles with strong
interactions (i.e. $b^e_3=0$) we can very easily derive the
following relations for $b^e_Y$ and $b^e_2$ which show their
dependence on the initial scale $M_0$
and on the scale $M_U$ where the couplings unify:
\begin{equation}\label{biextra}
\begin{split}
b^e_Y&=-\frac{2\pi}{t_U-t_0}\,\delta\alpha^{-1}_{3Y_0}+\delta b_{3Y}\\
b^e_2&=-\frac{2\pi}{t_U-t_0}\,\delta\alpha^{-1}_{32_0}+\delta
b_{32}\, .
\end{split}
\end{equation}
By eliminating $t_U-t_0$ we get the relation:
\begin{equation}\label{b12extra}
b^e_2=(b^e_Y-\delta
b_{3Y})\,\frac{\delta\alpha^{-1}_{32_0}}{\delta\alpha^{-1}_{3Y_0}}+\delta
b_{32}
\end{equation}

\begin{figure}[!t]
\centering
\begin{tabular}{cc}
\includegraphics[scale=0.8]{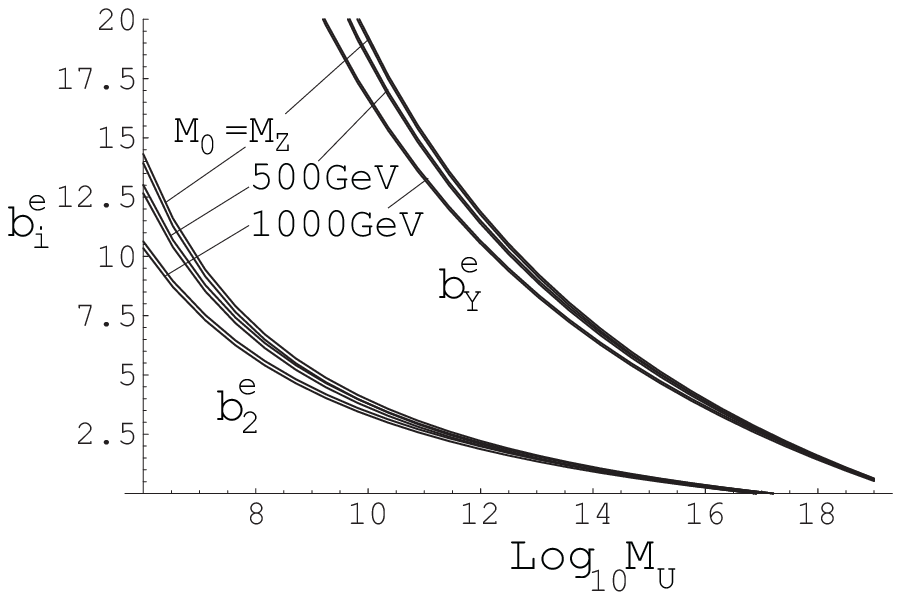}&
\includegraphics[scale=0.8]{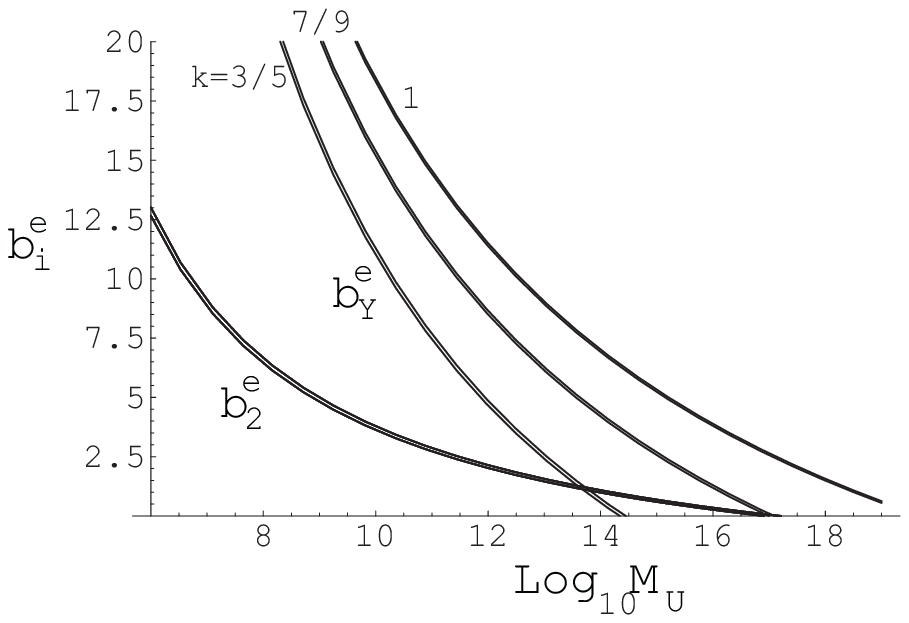}\\
(a)& (b)
\end{tabular}
\caption{The required $b^e_i$ coefficients from extra matter as a
function of the logarithm of the unification scale. (a) The
normalisation factor $k=1$ while the three lines for each coupling
corresponds to three choices of $M_0$ scale, namely: $M_Z$, 500
and 1000 GeV. (b) For $M_0=500$ GeV, the three lines of the
$b^e_Y$ corresponds to three values of the normalisation factor
$k=3/5$, $7/9$ and 1. The thickness of each line corresponds to
the experimental error to $\alpha_3$ (mainly) and $\sin^2\theta$
at the $M_Z$.} \label{FIG1}
\end{figure}

We point out that the scale $M_0$ is not necessarily $M_Z$ but the
scale where the extra matter start contributing to the running of
the gauge couplings. Therefore, $\alpha_{i_0}$ is the value at
this scale, running from $M_Z$ with the SM $b$ coefficients.

In Fig.\ref{FIG1}(a) we show the relations of Eq.(\ref{biextra})
keeping $M_0$ as a parameter: i.e. for each chosen scale of
unification $M_U$ in the $x$-axis, the $y$-axis gives the required
$b^e_2$ and $b^e_Y$ in order to achieve unification at that scale.
In each group the three lines correspond to the values $M_0=M_Z$,
500{\rm GeV} and {\rm 1TeV}. Finally, the thickness of the lines
corresponds to the errors of the experimental values of $\alpha_3$
(mainly) and $\sin^2\theta$ at $M_Z$.
%It is obvious that $b^e_2=0$ and $b^e_1=0$ is independent of $M_0$.
Since $b^e$ comes from matter contribution, it should be positive.
Therefore, from the figure we see that the maximum allowed $M_U$
is of the order of $10^{17}$ GeV, i.e. when $b^e_2$ becomes
negative. Also Fig.\ref{FIG1} shows the obvious result that the
lower the unification scale the richer the extra matter content
required to achieve unification. Finally, since the running of the
strong coupling is unaffected, $U(1)$ needs more matter content to
catch up the running of the $SU(2)$ coupling towards the strong
one.

\begin{figure}[!b]
\centering
\includegraphics[scale=0.7]{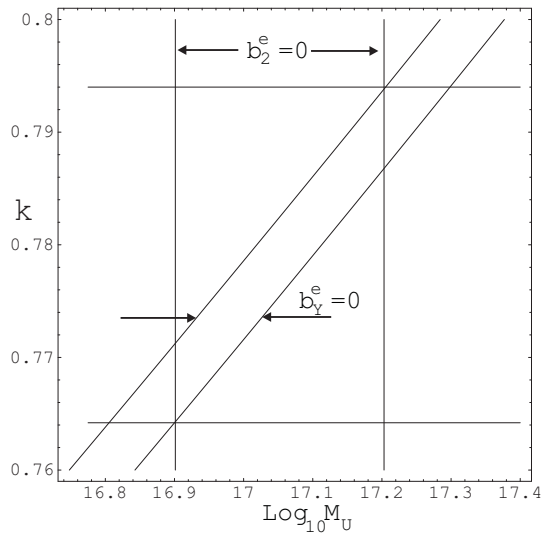}
\caption{Contours of $b^e_2=b^e_Y=0$ on the $(\log_{10}M_U,k)$
plane.} \label{KM_2}
\end{figure}

 Fig.\ref{FIG1}(b) is similar to Fig.\ref{FIG1} but we show the dependence on the $U(1)$ normalisation factor $k$
for the case $M_0=500$ GeV. We plot $b^e_2$ and $b^e_Y$ for
$k=3/5$, $7/9$ and 1 (for comparison). The  $k=7/9$ case is chosen
since it is a possible value where unification can be achieved
with no extra matter. The value $k=3/5$ corresponds of course to
the $SU(5)$ normalisation. We also see that, as $k$ is lowered,
$b^e_Y$ dictates the maximum allowed value of $M_U$ since it goes
to zero quicker than $b^e_2$. In Fig.\ref{KM_2} we present
the case of no extra matter by plotting contours of
$b^e_2=b^e_Y=0$ in the $(\log_{10}M_U,k)$ plane. The two vertical
lines (independent of $k$) defines the strip of $b^e_2=0$ while
the inclined strip corresponds to $b^e_Y=0$. The horizontal lines
show the allowed $k$ values for unification, $(0.764$ - $0.794)$,
while the corresponding value for $M_U$ is $\sim 10^{16.9-17.2}$
GeV.

\begin{figure}[!t]
\centering
\includegraphics[scale=0.7]{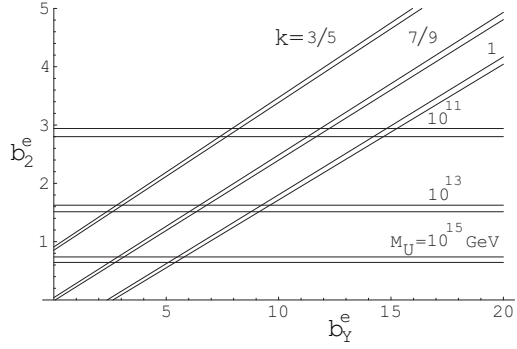}
\caption{The relation between $b^e_Y$ and $b^e_2$ for $M_0=500$
GeV and for three values of the normalisation factor $k=3/5$,
$7/9$ and $1$. We draw also three lines which correspond to
$M_U=10^{11}$, $10^{13}$ and $10^{15}$ GeV.} \label{KM_3}
\end{figure}

In Fig.\ref{KM_3} we show the relation between $b^e_Y$ and
$b^e_2$, Eq.(\ref{b12extra}), for $M_0=500$ GeV and for the three
values of $k=3/5$, $7/9$ and $1$. We indicate also three values of
the unification scale $M_U=10^{11}$, $10^{13}$ and $10^{15}$ GeV.

\begin{table}[!t]
\centering
\begin{tabular}{|c|c|c|c|c|c|}
\hline
$M_0$ (TeV)            & $b^e_Y$ & \multicolumn{2}{c|}{Scalars} &\multicolumn{2}{c|}{Fermions}  \\
                     &         & No & $Q$                     & No  &  $Q$                      \\
\hline
\multicolumn{6}{|l|}{$k=1$}\\
\hline
$M_Z\rightarrow$ 0.160 & 7/3     & 3  & 1                       & 2   & 1                         \\
\hline
0.450$\rightarrow$12    & 8/3     & 0 &  0                       & 4   & 1                       \\
                     &         & 2 &  2                       & 0   & 0                        \\
                     &         & 1 &  2                       & 2   & 1                        \\
                     &         & 4 &  1                       & 2   & 1                        \\
\hline
\multicolumn{6}{|l|}{$k=4/5$}\\
\hline
$M_Z\rightarrow 500$ &4/15 &   1    &   1   &   0   &   0\\
\hline
\end{tabular}
\caption{Regions of scale $M_0$ where unification is possible,
with $b^e_2=0$, and the indicated value for $b^e_Y$ for two values
of $k=1$ and $4/5$. The possible combinations of fermions and/or
scalars contributing the desired value to $b^e_Y$ are  are also
given.} \label{T1}
\end{table}

To make the analysis more comprehensive, our next step is to
further assume (still with $b^e_3=0$) that the extra matter
content has only $U(1)$ interactions and try to find the
multiplicity as well as the integer electric charges (or the
$U(1)$ quantum numbers) needed to have unification. In Table
\ref{T1} we give the regions where unification can be achieved as
well as the possible set of extra matter.  Of course, always
$M_U\sim 10^{17}$ GeV since $b^e_2=0$ and therefore the
unification scale is determined by the (MS) running of $\alpha_3$
and $\alpha_2$. For values of $k$ lower than $4/5$ the minimum
value that $b^e_Y$ could take is larger than the required $b^e_Y$,
and as we have seen in Fig.\ref{FIG1}(b), the lower the value of
$k$ the lower the required $b^e_Y$ to achieve unification, keeping
$b^e_2=0$.

\begin{table}[!b]
\centering
\begin{tabular}{|c|c|c|c|c|c|c|}
\hline
$b^e_Y$   &  $b^e_2$   & \multicolumn{2}{c|}{Scalars}  & \multicolumn{2}{c|}{Fermions}&  $M_U$ (GeV)  \\
          &            & No       & $Y$ charge        & No       & $Y$ charge       &                 \\
\hline
\multicolumn{7}{|l|}{$k=1$}\\
\hline
37/6      &   5/6      &      1   &    1/2            &      2   &       3/2        & (2.6-3.0)$10^{14}$\\
\hline
29/3      &   5/3      &      6   &    3/2            &      2   &       1/2        & (4.7-5.3)$10^{12}$\\
          &            &      2   &    5/2            &      4   &       1/2        &                   \\
\hline
79/6      &   5/2      &      7   &    1/2            &      4   &       3/2        & (2.4-2.6)$10^{11}$\\
\hline
89/6      &17/6        &      9   &    3/2            &      4   &       1/2        & (7.5-8.2)$10^{10}$\\
\hline
121/6     &   25/6     &      13  &    1/2            &      6   &       3/2        & (3.9-4.2)$10^{9}$\\
\hline
71/3      &   5        &      14  &    3/2            &      8   &       1/2        & (8.9-9.5)$10^{8}$\\
\hline
\multicolumn{7}{|l|}{$k=7/9$}\\
\hline
553/4     &   5/2      &      7   &    1/2            &       4  &        3/2       &  (2.3-2.7)$10^{11}$\\
\hline
623/4     &   17/6      &     9   &    3/2            &       4  &        1/2       &  (0.9-1.0)$10^{11}$\\
\hline
413/27    &   11/3      &    10   &    1/2            &      6   &        3/2       &  (0.9-1.0)$10^{10}$\\
          &             &    2    &    7/2            &     10   &        1/2       &                    \\
\hline
161/9     &   13/3      &    14   &    3/2            &      6   &        1/2       &  (2.5-2.8)$10^{9}$\\
\hline
553/27     &   5        &    14   &    1/2            &      8   &        3/2       &  (8.5-9.3)$10^{8}$\\
\hline
\multicolumn{7}{|l|}{$k=3/5$}\\
\hline
3/5       &   1         &    6    &    1/2            &      0   &        0         & (0.9-1.0)$10^{14}$\\
          &             &    2    &    1/2            &     2    &       1.2        &                   \\
\hline
22/5     &   2        &    8   &    1/2            &      2   &        3/2       &  (1.2-1.5)$10^{12}$\\
         &            &    4   &    3/2            &      4   &        1/2       &                      \\
\hline
129/10   &   25/6     &    13   &    3/2            &      6   &        1/2       &  (3.6-4.1)$10^{9}$\\
\hline
91/10     &   19/6    &    3   &    5/2            &      8   &        1/2       &  (3.1-4.7)$10^{10}$\\
\hline
69/5      &   13/3    &    14   &   3/2            &      6   &       1/2        &(2.3-2.7)$10^9$\\
\hline
157/10     &   29/6   &    13   &    1/2            &      8   &        3/2       &  (0.9-1.0)$10^{9}$\\
\hline
\end{tabular}
\caption{The required values of $b^e_2$ and $b^e_Y$ in order to
acheive unification of the couplings, for three values of $k$. We
assume that the extra matter forms $SU(2)$ doublets and is
colorless. The possible combinations of fermions and scalars with
the corresponding Y charges as well as the unification scale are
also given. We keep the number of extra multiplets under 15 and
$M_0=500$ GeV} \label{T2}
\end{table}

\begin{figure}[!b]
\centering
\includegraphics[scale=0.7]{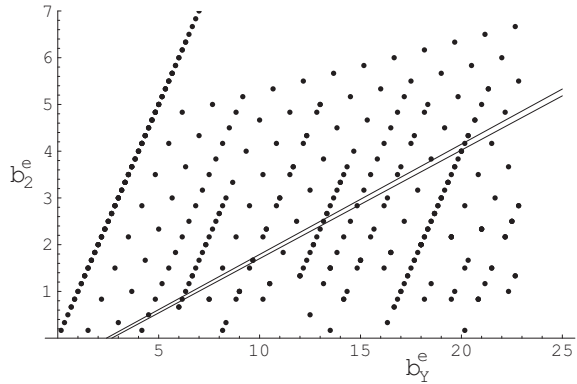}
\caption{The possible values of $b^e_1$ and $b^e_2$ are
represented by dots. The band corresponds to $M_0=500$ GeV. Only
the dots which are inside the band are acceptable for unification
($k=1$).} \label{FIG4}
\end{figure}

Now we turn to extra matter that has necessarily both $SU(2)$ {\it
and}  $U(1)$.
 Although the values that $b^e_1$ and $b^e_1$ can take is really great,
the number of acceptable cases stays relatively low. For example,
assuming that the extra particle, fermion and/or scalars, are in
the fundamental representation (doublet) with integer electric
charge and allowing up to 14 doublets for each kind, we get 2555
different combinations. Only 7 of them are leading to unification
(with $k=1$) if these extra particle starts to contribute at the
scale of $M_0={\rm 500 GeV}$ to the running. In Table \ref{T2} we
show several combinations of extra particles for two more values
of $k$. The unification scale runs from $10^{15}$ GeV down to
$10^8$ GeV. Of course, the lower the unification scale is the
reacher the new spectrum. In Fig.\ref{FIG4} we show the possible
combinations of $b^e_1$ and $b^e_2$ and the ones (inside the band)
that lead to unification (for $k=1$).

We can continue trying for higher representations. For example,
with just 2 scalar tetraplets with $Y=5/2$ we can achieve
unification at the scale $(2.8-3.1)10^{10}$ GeV for $M_0=500$ GeV
and $(4.2-4.5)10^{10}$ GeV for $M_0=1$ TeV.

In our next step we introduce colored extra matter. It consists of
one fermion multiplet with quantum numbers $(8,1)_0$ under
$SU(3)\times SU(2)\times U(1)$. i.e. a gluino-like particle. Of
course, the previous values of the $b^e_Y$ and $b^e_2$ correspond
now to the differences $b^e_Y-b^e_3$ and $b^e_2-b^e_3$ with $M_U$
unchanged. The results are shown in Table \ref{T3}. The connection
between the entries of Table \ref{T2} and Table \ref{T3} is
simple. Since we add a ``gluino'', we change $b^e_3$ by $+2$.
Therefore we should change also $b^e_Y$ and $b^e_2$ by the same
value in order to keep the differences the same. Thus, for $k=1$,
we can achieved unification with, for example, 6 extra fermions or
12 extra scalars or 4 fermions and 4 scalars, all doublets under
$SU(2)$ with $Y=1/2$. For $k\neq 1$ the required changes in
$b^e_2$ and $b^e_Y$ lead to small values of $k$ that do not fit to
the ones shown in Table\ref{T2}. Nevertheless, some interesting
minimal cases appear like the following: With no extra colour
particles, an extra $SU(2)$ scalar triplet with $Y=1$ is enough to
achieve unification around $10^{15}$ GeV when $k=2/3$. If we
assume an extra gluino-like particle, it suffices to add 3 more
scalars with the same quantum numbers.
\begin{table}[!h]
\centering

\begin{tabular}{|c|c|c|c|c|c|c|}
\hline
$\delta b^e_{Y3}$   &  $\delta b^e_{23}$   & \multicolumn{2}{c|}{Scalars}  & \multicolumn{2}{c|}{Fermions}&  $M_U$ (GeV)  \\
          &            & No       & $Y$ charge        & No       & $Y$ charge       &                 \\
          \hline
37/6      &   5/6      &      13  &    1/2            &      2   &       3/2        & (2.6-3.0)$10^{14}$\\
\hline
29/3      &   5/3      &      6   &    3/2            &      8   &       1/2        & (4.9-5.3)$10^{12}$\\
          &            &      2   &    5/2            &      10  &       1/2        &                   \\
\hline
71/3      &   5        &      14  &    3/2            &      14  &       1/2        & (8.9-9.5)$10^{8}$\\
\hline
\end{tabular}
\caption{The required values of the differences $b^e_2-b^e_3$ and
$b^e_Y-b^e_3$ in order to acheive unification of the couplings.
The extra colored matter consists of one fermion multiplet with
quantum numbers $(8,1)_0$ under $SU(3)\times SU(2)\times U(1)$
while the extra non colored matter consists of $SU(2)$ doublets.
The possible combinations of fermions and scalars of the non
colored matter with the corresponding Y charges as well as the
unification scale are also given. We keep the number of extra
multiplets under 15, $M_0=500$ GeV and $k=1$.} \label{T3}
\end{table}

Let us now consider a case which resembles the split supersymmetry
one, i.e. the extra matter contains one $(8,1)_0$ (gluino-like),
one $(1,3)_0$ (wino-like) and two $(1,2)_1$ (higgsinos-like). The
question is what extra matter, apart from those, are needed to
achieve unification. If we admit $U(1)$ only charged extra matter
(and $k=1$), with $M_0=500$ GeV, we need 3 scalars with $Y=3$ and
2 fermions with $Y=1$. If we choose $M_0=1$ TeV, there are
numerous solutions but the minimal one consists of four fermions
with $Y=2$. In both cases $M_U\sim 10^{13}$ GeV. If the extra
matter (again beyond the gluino-, wino- and higgsino-like) falls
in the fundamental representation of $SU(2)$, then the most
economical solution (again with $k=1$) is two scalars with $Y=1/2$
and two fermions with $Y=3/2$. The unification scale is $\sim
10^{14}$ GeV. If the extra matter falls in the triplet of $SU(2)$,
one scalar with $Y=4$ and two more wino-like multiplets are enough
to lead to unification at $M_U\sim 2\times 10^{10}$ GeV.

%\vspace{1cm}
\noindent
\section{\large{SO(10) Unification}}

There are various symmetry breaking patterns for $SO(10)$\cite{Lee:1994vp}
leading
to an intermediate group $G\subset SO(10)$, which  subsequently it
further breaks to  $SU(3)\times SU(2)\times U_Y(1)\subset G$. The
various possibilities for $G$ include (i) $G=SU(5)$, (ii) the
Pati-Salam $G=SU(4)\times SU(2)\times SU(2)$  (iii) the flipped
$G=SU(5)\times U(1)$, iv)$G=SU(3)\times U(1)\times SU(2)\times
SU(2)$ e.tc. Here we will consider the cases of the Pati-Salam and
flipped $SU(5)$ (ii) and (iii), respectively.
\enlargethispage{-3\baselineskip}

\vspace{0.5cm} \noindent
{\bf The Pati-Salam $SU(4)\times SU(2)_L\times SU(2)_R$ GUT model}\\

We will consider here the possibility of unification through the
Pati-Salam GUT model $SU(4)\times SU(2)_L\times SU(2)_R$. The
scalar sector contains a 210, a 126 and a 10 of $SO(10)$. The 210
breaks $SO(10)\to SU(4)\times SU(2)_L\times SU(2)_R$, the 126
breaks $SU(4)\times SU(2)_L\times SU(2)_R$ to $SU(3)_C\times
SU(2)_L\times U(1)_R\times U(1)_C$, whereas the 10 further breaks
to the SM group $SU(3)_C\times SU(2)\times U(1)_Y$. The symmetry
pattern is then
\[
\begin{split}
SO(10)\!\stackrel{210}{\longrightarrow}\! SU(4)\!\times
\!SU(2)_L\!\times\! SU(2)_R &\stackrel{126}{\longrightarrow}
SU(3)_C\!\times\! SU(2)_L\times\! U(1)_R\!\times\! U(1)_C
\stackrel{10}{\longrightarrow}\\
SU(3)_C\!\times\! S&U(2)\!\times \!U(1)_Y
\end{split}
\]
At the GUT scale we have the relation:
\begin{equation}
\alpha_4=\alpha_3,\quad\alpha_{2L}=\alpha_2,\quad\alpha_Y^{-1}=3/5\,\alpha_{2R}^{-1}+2/5\,\alpha_4^{-1}
\end{equation}
while the $U(1)_Y$ quantum numbers are related to the
corresponding ones in $SU(4)$ and $SU(2)_R$ through the relation:
\begin{equation}
Y=\sqrt{3/5}\,Y_{2R}+\sqrt{2/5}\,Y_4
\end{equation}
The SM particles (quarks and leptons plus a right handed neutrino)
are in the $(4,2,1)$ and $(\bar{4},1,2)$ representations of
$SU(4)\times SU(2)_L\times SU(2)_R$ (these two multiplets make the
16 representation of $SO(10)$). The required Higgs to break the
Pati-Salam group down to the SM are in the
%$(4,1,2)$ and $(\bar{4},1,2)$
$(\bar{10},1,3)$ representations while the SM Higgs is in the
$(1,2,2)$ representation.

The energy scales we are using are the following:
\begin{itemize}
\item $M_U$, where $\alpha_4=\alpha_{2L}=\alpha_{2R}$, \item
$M_G$, where the Pati-Salam model breaks to the SM. In the region
between $M_U$ and $M_G$ we have the minimal content of the
Pati-Salam group  consisting of SM particles in $(4,2,1)$ and
$(\bar{4},1,2)$, Higgs fields in $(\bar{10},1,3)$ and $(1,2,2)$
plus some extra fermionic matter, \item Below $M_G$, down to some
scale $M_N$, we have the SM particles in $(4,2,1)$ and the
$(\bar{4},1,2)$, $(1,2,2)$ Higgs,
 plus some extra fermionic matter and
\item Below $M_N$ we have only the SM content.
\end{itemize}

We assume that all extra fermionic matter  come from the 10, 16
and 45 representations of $SO(10)$. We have 16 cases which are
shown in Table \ref{T4}. In each case, the first line gives the quantum numbers under the
corresponding group. The second line gives the contribution to the
SM $\beta$-functions and to the Pati-Salam model for one
multiplet.
\begin{table}[!t]
\[
\begin{array}{|c|ccc|}\hline
CASE   & SU(3)\times SU(2)\times U(1)_Y  & SU(4)\times SU(2)_L\times SU(2)_R & SO(10)\\
\hline
1      & \mathbf{(3,1,\sqrt{4/15})}               &    \mathbf{(4,1,2)}                        & \mathbf{16}\\
       & (b_3,b_2,b_1)=(1/3,0,8/15)                    &  (b_4,b_{2L},b_{2R})=  (2/3,0,4/3)                    &\\
\hline
2      & \mathbf{(3,1,-\sqrt{1/15})}              &    \mathbf{(4,1,2)}                        &\mathbf{16}\\
       & (1/3,0,2/15                     &     (2/3,0,4/3)                   &\\
\hline
3      & \mathbf{(\stackrel{(-)}{3},1,\mp\sqrt{1/15})} & \mathbf{(6,1,1)}                      &\mathbf{10}\\
       & (1/3,0,2/15)                    &     (2/3,0,0)                     &\\
\hline
4      & \mathbf{(\stackrel{(-)}{3},1,\pm\sqrt{2/15})} & \mathbf{(15,1,1)}                     & \mathbf{45}\\
       & (1/3,0,4/15)                    &     (8/3,0,0)                     &\\
\hline
5      & \mathbf{(\stackrel{(-)}{3},2,\pm\sqrt{1/60})} & \mathbf{(6,2,2)}                      & \mathbf{45}\\
       & (2/3,1,1/15)                     &   (8/3,4,4)                      &\\
\hline
6      & \mathbf{(\stackrel{(-)}{3},1,\pm 5\sqrt{60})} & \mathbf{(6,2,2)}                      &\mathbf{45}\\
       & (2/3,1,5/3)                     &     (8/3,4,4)                     &\\
\hline
7      & \mathbf{(1,1,\pm\sqrt{3/5})}             & \mathbf{(1,1,3)}                           & \mathbf{45}\\
       & (0,0,2/5)                       &   (0,0,4/3)                       &\\
\hline
8      & \mathbf{(1,3,0)}                         & \mathbf{(1,3,1)}                           & \mathbf{45}\\
       & (0,4/3,0)                    &   (0,4/3,0)                       &\\
\hline
9      & \mathbf{(8,1,0)}                         & \mathbf{(15,1,1)}                          & \mathbf{45}\\
       & (2,0,0)                         &  (8/3,0,0)                        &\\
\hline
10     & \mathbf{(1,1,0)}                         & \mathbf{(15,1,1)}                          & \mathbf{45}\\
       & (0,0,0)                         &   (8/3,0,0)                       &\\
\hline
11     & \mathbf{(1,1,0)}                         & \mathbf{(4,1,2)}                           & \mathbf{16}\\
       & (0,0,0)                         &   (2/3,0,4/3)                     &\\
\hline
12     & \mathbf{(1,1,0)}                         & \mathbf{(1,1,3)}                           & \mathbf{45}\\
       & (0,0,0)                         &   (0,0,4/3)                       &\\
\hline
13     & \mathbf{(1,2,\pm\sqrt{3/20})}   & \mathbf{(1,2,2)}                                    &\mathbf{10}\\
       & (0,1/3,1/5)                         &   (0,1/3,1/3)                 &\\
\hline
14     & \mathbf{(\bar{3},2,-\sqrt{1/60})} & \mathbf{(\bar{4},2,1)}                           & \mathbf{16}\\
       & (2/3,1,1/15)                         &   (2/3,4/3,0)                       &\\
\hline
15     & \mathbf{(1,2,\sqrt{3/20})}           & \mathbf{(\bar{4},2,1)}                           & \mathbf{16}\\
       & (0,1/3,1/5)                         &   (2/3,4/3,0)                       &\\
\hline
16     & \mathbf{(1,1,-\sqrt{3/5})}         & \mathbf{(4,1,2)}                                & \mathbf{16}\\
       & (0,0,2/5)                         &   (2/3,0,4/3)                       &\\
\hline
\end{array}
\]
\caption{The 16 possible cases of $SU(4)\times SU(2)_L\times SU(2)_R$ which fall in the 10, 16 and 45 multiplets of $SO(10)$.
The corresponding contribution to the $\beta$-functions are also given for each case in the second line.}
\label{T4}
\end{table}

The number of free parameters are great: $M_N$, $M_G$ and the number of each multiplet. We are going
to limit this parameter space, choosing first, to make the running independent of the scale $M_N$. This could happen
if the contribution of the extra matter to the SM $\beta$-functions are equal (since the unification depends only
on the differences of the contributions).
%Notice that c.12 is very convenient since it affects only $SU(2)_R$ running and could play a ``ballader" role in achieving unification.
%Therefore, the range for $M_G$ and $M_U$ is large.
For example, if the contribution to the SM $\beta$-function coefficients from extra matter is 2/3, choosing
$M_G\sim 6.3\times 10^{6}$ GeV, we can achieve unification at $M_U\sim 4\times 10^{17}$ GeV .
A minimal extra matter content for that case is
2 multiplets from case 2 and from case 13 of Table \ref{T4}.
Notice that in this standard (non-SUSY) version of the PS model,
we are safe against proton decay and we should only take care of a high
$M_U$ energy scale \cite{Lee:1994vp}.

\begin{figure}[!t]
\centering
\includegraphics[scale=0.9]{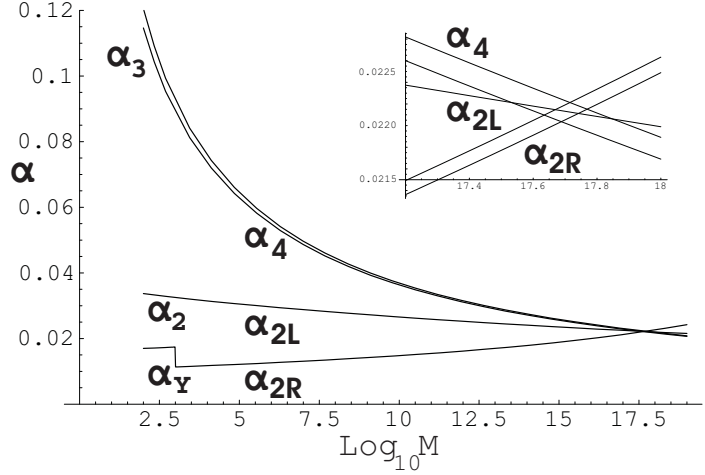}
\caption{ The coupling running for the PS model. $M_N=M_G=1$ GeV.
The extra content above $M_G$ is 2 multiplets from case 2, 1 from case 7 and 1 from case 13 of Table \ref{T5}.}
\label{FIG5}
\end{figure}

Another idea that we have checked is the possibility that $M_N=M_G$, i.e. the extra
matter is only in the energy region of the Pati-Salam model (not requiring necessarily equality of SM extra contributions).
We have found that if $M_N=M_G=1$ TeV we can achieve unification at $M_U\sim 5\times 10^{17}$ GeV.
In Table \ref{T5} we show the lowest possible extra content in the region between $M_G$ and $M_U$.
In   Fig.\ref{FIG5} we plot the corresponding running of the couplings.
If we rase the 1 TeV scale to 5 TeV, then we get unification at $M_U\sim 6\times 10^{14}$ GeV.

\begin{table}
\[
\begin{array}{|c|c|c|c|c|c|c|c|}
\hline
(4,1,2)  & (6,1,1)  &  (15,1,1)  &  (6,2,2)  &  (1,1,3)& (1,3,1)&(1,2,2)&(\bar{4},2,1) \\
\hline
   2     &    0     &     0      &     0     &     1   &   0    &   1    &     0\\
   \hline
\end{array}
\]
\caption{The lowest possible combinations of Pati-Salam multiplets for $M_N=M_G=1$ TeV
and $M_U\sim 5\times 10^{17}$ GeV.}
\label{T5}
\end{table}

\vspace{1cm} \noindent
{\bf The flipped $SU(5)$ GUT model}\\
The $SU(5)\times U(1)$ group is another possible GUT model,
subgroup of $SO(10)$. At the GUT scale we have the relations:
\begin{equation}
\alpha_5=\alpha_3=\alpha_2,\quad\alpha_Y^{-1}=1/25\,\alpha_5^{-1}+24/25\,\alpha_1^{-1}
\end{equation}
The $U(1)_Y$ quantum numbers are related to the corresponding
$U(1)$'s in $SU(5)$ and the $U(1)$ by the relation
\begin{equation}
Y=-\frac{1}{5}\,Y_5+\frac{\sqrt{24}}{5}\,Y_1
\end{equation}
The matter content lie in the 5$^*$, 10 and 1 of $SU(5)$ (again
these three multiplets make the 16 of $SO(10)$) and the higgs are
in the 10 and 5$^*$. We allow extra matter (either in $SU(5)\times
U(1)$ or in the SM region) that can be found in the 10, 16 and 45
of $SO(10)$. There are 12 cases shown in Table \ref{T6}.

\begin{table}[!t]
\[
\begin{array}{|c|ccc|}
\hline
CASE   & SU(3)\times SU(2)\times U(1)_Y  & SU(5)\times U(1)                  & SO(10)\\
\hline
1      & \mathbf{(3,1,-\sqrt{1/15})}     &    \mathbf{(5,-\sqrt{1/10})}      & \mathbf{10}\\
       & (b_3,b_2,b_Y)=(1/3,0,2/15)      &    (b_5,b_1)=(1/3,1/3)                      &\\
\hline
2      & \mathbf{(3,1,-\sqrt{1/15})}     &    \mathbf{(10,\sqrt{1/40})}      &\mathbf{16}\\
       & (1/3,0,2/15)                    &     (1,1/6)                       &\\
\hline
3      & \mathbf{(1,2,-\sqrt{3/20})}     & \mathbf{(5,-\sqrt{1/10})}         &\mathbf{10}\\
       & (0,1/3,1/5)                    &     (1/3,1/3)                      &\\
\hline
4      & \mathbf{(1,2,-\sqrt{3/20})}     & \mathbf{(\stackrel{-}{5},-\sqrt{9/40})}         &\mathbf{16}\\
       & (0,1/3,1/5)                     &     (1/3,3/4)                     &\\
\hline
5      & \mathbf{(3,2,\sqrt{1/60})}       & \mathbf{(10,\sqrt{1/40})}                      & \mathbf{16}\\
       & (2/3,1,1/15)                     &   (1,1/6)                      &\\
\hline
6      & \mathbf{(3,2,\sqrt{1/60})}        & \mathbf{(24,0)}                      &\mathbf{45}\\
       & (2/3,1,1/15)                     &     (10/3,0)                     &\\
\hline
7      & \mathbf{(\stackrel{-}{3},1,-\sqrt{4/15})} & \mathbf{(\stackrel{-}{5},-\sqrt{9/40})}                           & \mathbf{16}\\
       & (1/3,0,8/15)                       &   (1/3,3/4)                       &\\
\hline
8      & \mathbf{(\stackrel{-}{3},1,-\sqrt{4/15})}                         & \mathbf{(10,-\sqrt{2/5})}                           & \mathbf{45}\\
       & (1/3,0,8/15)                    &   (1,8/3)                       &\\
\hline
9      & \mathbf{(1,1,\sqrt{3/5})}         & \mathbf{(1,\sqrt{5/8})}                          & \mathbf{16}\\
       & (0,0,2/5)                         &  (0,5/12)                        &\\
\hline
10     & \mathbf{(1,1,\sqrt{3/5})}         & \mathbf{(10,-\sqrt{2/5})}                          & \mathbf{45}\\
       & (0,0,2/5)                         &   (1,8/3)                       &\\
\hline
11     & \mathbf{(8,1,0)}                & \mathbf{(24,0)}                           & \mathbf{45}\\
       & (2,0,0)                         &   (10/3,0)                     &\\
\hline
12     & \mathbf{(3,2,\sqrt{5/12})}          & \mathbf{(10,-\sqrt{2/5})}                           & \mathbf{45}\\
       & (2/3,1,5/3)                         &   (1,8/3)                       &\\
\hline
13     & \mathbf{(1,1,0)}          & \mathbf{(10,\sqrt{1/40})}                           & \mathbf{16}\\
       & (0,0,0)                         &   (1,1/6)                       &\\
\hline
14     & \mathbf{(1,1,0)}          & \mathbf{(24,0)}                           & \mathbf{45}\\
       & (0,0,0)                   &   (10/3,0)                       &\\
\hline
\end{array}
\]
\caption{The 14 possible cases of $SU(5)\times U(1)$ which fall in
the 10, 16 and 45 multiplets of $SO(10)$. The corresponding
contribution to the $\beta$-functions are also given.} \label{T6}
\end{table}

\begin{figure}[!t]
\centering
\includegraphics[scale=0.9]{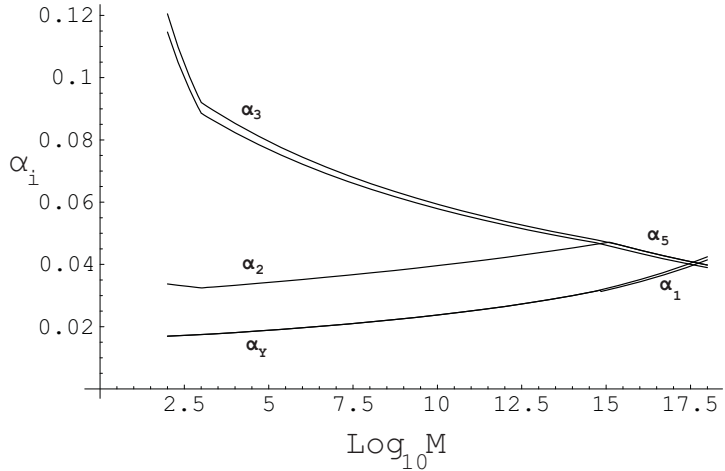}
\caption{ The coupling runnings for the flipped $SU(5)$ model. The
new SM multiplets appear at the 1 TeV scale. For the specific
choice of extra multiplets, (0,0,2,2,2,2,0,0,2,0,1,0,0,0), $M_G\sim
10^{15}$ GeV while $M_U\sim 4\, 10^{17}$ GeV. }
\label{FIG6}
\end{figure}

In this model the grand unification scale $M_G$ is where the
$SU(3)$ and the $SU(2)$ couplings meet. Then the couplings of
$SU(5)$ and $U(1)$ run and their meeting point indicates the
unification scale $M_U$.
Asking the number of extra multiplets for each case to be
at most 2 (the lowest possible value), and requiring that $M_G\geq
10^{14}$ GeV (to avoid fast proton decay) and $M_U\leq 10^{18}$ GeV,
we get numerous solutions.
However, they all fall in two categories:
(i) both $M_G$ and $M_U$ are
in the scale of $10^{17}$ GeV and
(ii) $M_G\sim (0.7-1)10^{15}$ GeV and
$M_U$ could be either $(0.2-4)10^{16}$ GeV or $(2-4)10^{17}$ GeV. In Fig. \ref{FIG6} we show
the running of the couplings for the case
(0,0,2,2,2,2,0,0,2,0,1,0,0,0), where the  numbers correspond to the number of species
from each of the fourteen cases in Table \ref{T6}.  We have assumed, as before, that
the extra SM multiplets appear at the scale of 1 TeV.

%\vspace{1cm}

\section{\large Conclusions}
%\vspace{-.1cm}

We had performed a wide (1-loop analysis) answering the question:
``How much and what kind of extra matter should we add to the SM
content to achieve unification?". Forgetting, in the first step,
any covering group, we found numerous solutions with $U(1)$ only
charged extra matter (therefore the unification scale stays in the
order of $M_U=10^{16}$ GeV), or with $SU(2)$ charged extra matter
($M_U$ starts to get lower) as well as for the case of gluino-like
extra matter. We also examine the case of various $U(1)_Y$
normalisations as it happens in GUT theories on orbifolds
\cite{Kawamura} and in their
deconstructions~\cite{Arkani-Hamed:2001ca} and D-brane derived
models \cite{Blumenhagen:2005mu}. This  corresponds to unification
of the SM into a more fundamental theory, like string theory,
without any GUT intermediate step. Such gauge coupling unification
has been considered before \cite{munoz},
\cite{Ibanez:1993bd},\cite{Dienes:1996du},\cite{Barger:2005gn}
Then we have identified the SM extra matter as coming from the
breaking of $SO(10)$
\cite{Lee:1994vp},\cite{Calmet}
 either through the Pati-Salam group
$SU(4)\times SU(2)_L\times SU(2)_R$ or flipped $SU(5)$. We have not discussed some other
breaking of $SO(10)$, which are triggered by Higgs scalars in other $SO(10)$
representations~\cite{Lazarides:1980cc}.
For  both
cases we present, unification can be achieved by a rather minimal extra
content. This is reminiscent of the work in~\cite{Giudice:2004tc},
where however, a one-step unification has been followed and the
$SU(3)\times SU(2)\times U_Y(1)$ is unified to $SU(5)$. In this
case, proton stability requires the existence of coloured
particles in the electroweak scale. These particles, if stable,
could be bound on nuclei, giving rise to anomalous heavy isotopes.
There are various searches of such heavy isotopes using deep sea
water, which put bounds on the concentration of stable charged
particles of less than $10^{17}$ for mass up to $1600 ~
m_p$~\cite{YY}.
However, in the case of a two-step unification, i.e., in the case
of an $SO(10)$ unification with a partial intermediate unification
of Pati-Salam or flipped $SU(5)$ unification, unification may
achieved without introducing necessarily coloured particles. In both cases,
there are stable dark matter candidates, while the splitting of
irreps (the double-triplet splitting in the case of $SU(5)$) is
the same for both one-step and two-step unification.

\vspace{1cm}
\noindent
The project  is co - funded by the European Social Fund (75\%)
and National Resources (25\%)  - (EPEAEK II) -PYTHAGORAS.
The work of NDT is partially supported by
the MRTN-CT-2004-503369 European Network.

%\newpage


\begin{thebibliography}{99}
\bibitem{AD}
N.~Arkani-Hamed and S.~Dimopoulos,
%``Supersymmetric unification without low energy supersymmetry and signatures
%for fine-tuning at the LHC,''
arXiv:hep-th/0405159.
%%CITATION = HEP-TH 0405159;%%

\bibitem{Giudice:2004tc}
G.~F.~Giudice and A.~Romanino,
%``Split supersymmetry,''
Nucl.\ Phys.\ B {\bf 699}, 65 (2004) [Erratum-ibid.\ B {\bf 706},
65 (2005)] [arXiv:hep-ph/0406088].
%%CITATION = HEP-PH 0406088;%

\bibitem{AD1}
N.~Arkani-Hamed, S.~Dimopoulos, G.~F.~Giudice and A.~Romanino,
%``Aspects of split supersymmetry,''
Nucl.\ Phys.\ B {\bf 709}, 3 (2005) [arXiv:hep-ph/0409232].
%%CITATION = HEP-PH 0409232;%%

\bibitem{AnD}
I.~Antoniadis and S.~Dimopoulos,
%``Splitting supersymmetry in string theory,''
Nucl.\ Phys.\ B {\bf 715}, 120 (2005) [arXiv:hep-th/0411032].
%%CITATION = HEP-TH 0411032;%%

\bibitem{KN}
B.~Kors and P.~Nath,
%``Hierarchically split supersymmetry with Fayet-Iliopoulos D-terms in string
%theory,''
Nucl.\ Phys.\ B {\bf 711}, 112 (2005) [arXiv:hep-th/0411201].
%%CITATION = HEP-TH 0411201;%%

%\cite{Pati:1974yy}
\bibitem{Pati:1974yy}
  J.~C.~Pati and A.~Salam,
  %``Lepton Number As The Fourth Color,''
  Phys.\ Rev.\ D {\bf 10}, 275 (1974).
  %%CITATION = PHRVA,D10,275;%%


%\cite{Barr:1981qv},\cite{Derendinger:1983aj},\cite{Antoniadis:1987dx}
\bibitem{Barr:1981qv}
  S.~M.~Barr,
  %``A New Symmetry Breaking Pattern For SO(10) And Proton Decay,''
  Phys.\ Lett.\ B {\bf 112}, 219 (1982).
  %%CITATION = PHLTA,B112,219;%%

  %\cite{Derendinger:1983aj}
\bibitem{Derendinger:1983aj}
  J.~P.~Derendinger, J.~E.~Kim and D.~V.~Nanopoulos,
  %``Anti - SU(5),''
  Phys.\ Lett.\ B {\bf 139}, 170 (1984).
  %%CITATION = PHLTA,B139,170;%%

%\cite{Antoniadis:1987dx}
\bibitem{Antoniadis:1987dx}
  I.~Antoniadis, J.~R.~Ellis, J.~S.~Hagelin and D.~V.~Nanopoulos,
  %``Supersymmetric Flipped SU(5) Revitalized,''
  Phys.\ Lett.\ B {\bf 194}, 231 (1987).
  %%CITATION = PHLTA,B194,231;%%


\bibitem{Hung}
P.~Q.~Hung,
  %``Minimal SU(5) resuscitated by long-lived quarks and leptons,''
  Phys.\ Rev.\ Lett.\  {\bf 80}, 3000 (1998)
  [arXiv:hep-ph/9712338].


%\bibitem{Anchordoqui:2004bd}
%L.~Anchordoqui, H.~Goldberg and C.~Nunez,
%``Probing split supersymmetry with cosmic rays,''
%Phys.\ Rev.\ D {\bf 71}, 065014 (2005) [arXiv:hep-ph/0408284].
%%CITATION = HEP-PH 0408284;%%




\bibitem{Lee:1994vp}
D.~G.~Lee, R.~N.~Mohapatra, M.~K.~Parida and M.~Rani,
%``Predictions for proton lifetime in minimal nonsupersymmetric SO(10) models:
%An update,''
Phys.\ Rev.\ D {\bf 51}, 229 (1995)
[arXiv:hep-ph/9404238].
%%CITATION = HEP-PH 9404238;%%.




\bibitem{Kawamura}
Y.~Kawamura, Prog.\ Theor.\ Phys.\  {\bf 103}, 613 (2000); G.
Altarelli and F. Feruglio, Phys.\ Lett.\ B {\bf 511}, 257 (2001);
L. Hall and Y. Nomura, Phys.\ Rev.\ D {\bf 64}, 055003 (2001); A.
Hebecker and J. March-Russell, Nucl.\ Phys.\ B {\bf 613}, 3
(2001);
 T. Li, Phys.\ Lett.\ B {\bf 520}, 377 (2001);
 Nucl.\ Phys.\ B {\bf 633}, 83 (2002)




%
\bibitem{Arkani-Hamed:2001ca}
N.~Arkani-Hamed, A.~G.~Cohen and H.~Georgi,
%``(De)constructing dimensions,''
Phys.\ Rev.\ Lett.\  {\bf 86}, 4757 (2001);
%[arXiv:hep-th/0104005].
%%CITATION = HEP-TH 0104005;%%
C.~T.~Hill, S.~Pokorski and J.~Wang,
%``Gauge invariant effective Lagrangian for Kaluza-Klein modes,''
Phys.\ Rev.\ D {\bf 64}, 105005 (2001).
%[arXiv:hep-th/0104035].
%%CITATION = HEP-TH 0104035;%%





%
\bibitem{Blumenhagen:2005mu}
  R.~Blumenhagen, M.~Cvetic, P.~Langacker and G.~Shiu,
  %``Toward realistic intersecting D-brane models,''
  arXiv:hep-th/0502005.
  %%CITATION = HEP-TH 0502005;%%

\bibitem{munoz} J.A. Casas and C. Munoz,
Phys.\ Lett.\ B {\bf 214} 543 (1988).

%
\bibitem{Ibanez:1993bd}
L.~E.~Ibanez,
%``Gauge coupling unification: Strings versus SUSY GUTs,''
Phys.\ Lett.\ B {\bf 318}, 73 (1993) [arXiv:hep-ph/9308365].
%%CITATION = HEP-PH 9308365;%%

%
\bibitem{Dienes:1996du}
K.~R.~Dienes,
%``String Theory and the Path to Unification: A Review of Recent Developments,''
Phys.\ Rept.\  {\bf 287}, 447 (1997) [arXiv:hep-th/9602045].
%%CITATION = HEP-TH 9602045;%%



%
\bibitem{Barger:2005gn}
V.~Barger, J.~Jiang, P.~Langacker and T.~Li,
%``Gauge coupling unification in the standard model,''
arXiv:hep-ph/0503226.
%%CITATION = HEP-PH 0503226;%%


\bibitem{Calmet}
X.~Calmet,
  %``Minimal grand unification model in an anthropic landscape,''
  Eur.\ Phys.\ J.\ C {\bf 41}, 245 (2005)
  [arXiv:hep-ph/0406314].



%\cite{Lazarides:1980cc}
\bibitem{Lazarides:1980cc}
  G.~Lazarides, M.~Magg and Q.~Shafi,
  %``Phase Transitions And Magnetic Monopoles In SO(10),''
  Phys.\ Lett.\ B {\bf 97}, 87 (1980);
  %%CITATION = PHLTA,B97,87;%%
  D.~Chang and A.~Kumar,
  %``Symmetry Breaking Of SO(10) By 210-Dimensional Higgs Boson And The Michel's
  %Conjecture,''
  Phys.\ Rev.\ D {\bf 33}, 2695 (1986);
  %%CITATION = PHRVA,D33,2695;%%
  X.~G.~He and S.~Meljanac,
  %``Stability Of Spontaneous Symmetry Breaking In A Class Of SO(10) Models,''
  Phys.\ Rev.\ D {\bf 40}, 2098 (1989);
  J.~Basecq, S.~Meljanac and L.~O'Raifeartaigh,
  %``Stability Of Spontaneous Symmetry Breaking In A Class Of SO(10) Models,''
  Phys.\ Rev.\ D {\bf 39}, 3110 (1989);
  %%CITATION = PHRVA,D39,3110;%%
  %\cite{Kaymakcalan:1985us}
  O.~Kaymakcalan, L.~Michel, K.~C.~Wali, W.~D.~McGlinn and L.~O'Raifeartaigh,
  %``Absolute Minima Of A SO(10) Invariant Higgs Potential,''
  Nucl.\ Phys.\ B {\bf 267}, 203 (1986).
  %%CITATION = NUPHA,B267,203;%%




\bibitem{YY}
T. Yamagata, Y. Takamori, and H. Utsunomiya, Phys. \ Rev. \  D
{\bf 47}, 1231 (1993).

\bibitem{JJ}
D. Javorsek, D. Elmore1, E. Fischbach1, D. Granger, T. Miller, D.
Oliver, and V. Teplitz, Phys. \ Rev.\  Lett. \ {\bf 87}, 231804
(2001).

\end{thebibliography}
\end{document}